\documentstyle[aps, multicol, epsfig]{revtex}
\def\bea{\begin{eqnarray}}
\def\eea{\end{eqnarray}}
\def\be{\begin{equation}}
\def\ee{\end{equation}}
\begin{document}


\author{Bogdan Damski$^{1,2}$, Krzysztof Sacha$^1$, and Jakub Zakrzewski$^1$}
\address{
$^1$ Instytut Fizyki imienia Mariana Smoluchowskiego,
  Uniwersytet Jagiello\'nski,\\
 ulica Reymonta 4, PL-30-059 Krak\'ow, Poland \\
$^2$ Institut f\"ur Theoretische Physik, Universit\"at Hannover, D-30167
Hannover, Germany\\}
\title{Stirring Bose-Einstein condensate}
\date{\today}
\maketitle
\begin{abstract}
By shining a tightly focused laser light on the condensate
and moving the center of the beam along the spiral line one may
stir the condensate and create vortices.
It is shown that one can  induce rotation of the condensate 
in the direction opposite to the direction of the stirring. 
\end{abstract}

\begin{multicols}{2}

When a spoon (stick, or another object of a similar shape) is used to stir
a liquid, the latter rotates in the direction induced by the stirring
object. Is it possible to make the liquid rotating counter-clockwise
while stirring it clockwise? The aim of this paper is to show that
this counter-intuitive scenario may be realized in a quantum fluid,
or more precisely a Bose-Einstein condensate (BEC) when stirring it
 with the help of a tightly focused laser beam. 

Creation of vortices in a BEC and study of their properties has been a subject of
quite intensive research for last couple of years (an extensive list of
references may be found in \cite{httpBEC}). Let us mention that 
 vortices have been created in the BEC 
 experimentally using various methods.
 Paris group \cite{Madison00}
 used the rotating anisotropic potential (created by a 
detuned broad laser beam) to make a direct analog of the rotating bucket
  experiments \cite{Leggett92}. The formation of a vortex is then the result 
  of dynamical instabilities appearing in the course of the experiment 
  \cite{castin}. 
  Similar method was used by Ketterle group \cite{abo-shaeer2001}. Boulder
  group 
   \cite{Matthews99} created vortices in two component condensate, where 
  one fraction was made to rotate with respect to
  the other by means of the phase engineering technique.
  The latter technique   
  attempts to create directly the desired vortex
  state in the condensate.

  Various ``stirring'' propositions
 have been  discussed theoretically \cite{castin,caradoc99} for
 creation of vortices. In particular \cite{caradoc99} used a 
 localized potential moving on a circular path (with an appropriate
 smooth turn-on and turn-off of the stirrer). Such a stirring produces
 the condensate state which may be approximately described as a time
 dependent
 combination of the ground state and the vortex state. As time evolves 
 the system undergoes a generalized Rabi oscillation between the ground
 state and the vortex state. 

Similar in spirit is our recent proposition
for creation of vortices in a BEC
\cite{damski01}. 
It relies on an appropriate deformation 
of a harmonic trapping potential by means of an additional, tightly focused
 laser beam. 
The beam approaches the center of the trap moving along a spiral line. The
effective interaction of the  detuned laser beam 
with atoms results in an additional effective potential seen by the 
atomic external degrees of freedom. 
Neglecting the interaction between atoms, the effective two-dimensional 
Hamiltonian, in the frame rotating  with the center of the laser beam,
  reads
\begin{eqnarray}
\label{ham}
\hat{H} & = & -\frac{1}{2}\frac{\partial^2}{\partial x^2}-
\frac{1}{2}\frac{\partial^2}{\partial y^2}-\Omega \hat{L}_z 
+ \frac{x^2 + y^2}{2} \cr
&& 
-u_0\arctan(|x_0|) \exp\left(-\frac{(x - x_0)^2 + y^2}{2 \sigma^2} \right),
\end{eqnarray}
where $\Omega$ is the frequency of the rotation of the laser beam around the
center of the trap while $u_0>0$, $\sigma$ and $x_0$ stand for 
the parameters of the beam. 
In Eq. (\ref{ham})  and in the following we work in units defined by
the harmonic trap.
Changing $x_0$ from an initial negative  value to zero,  according to
$x_0(t)=x_0(0)+vt$,  corresponds to the motion of the laser beam 
along a spiral line in
the laboratory frame. 
When the beam reaches the center of the trap its intensity
is reduced to zero [see $\arctan(|x_0|)$ function in Eq.~(\ref{ham})] 
and we end up with the harmonic trapping potential only (for details
of the method see ~\cite{damski01}). 

\begin{figure}
\centering
{
\includegraphics[width=8.0cm, clip=true]{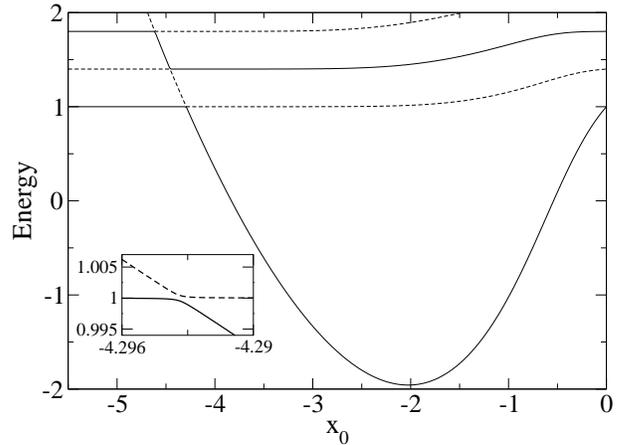}
}
\caption{Energy levels of the Hamiltonian ({\protect \ref{ham}}) as a function
of $x_0$ for $u_0=16$, $\sigma=0.2$ and $\Omega=0.6$.
The energy levels for $x_0=0$ correspond to $L_z=0$,
$L_z=1$ and $L_z=2$ (from bottom to top). Note very narrow anti-crossing
structures between the neighboring energy levels.  The inset shows the
anti-crossing between the ground and first excited levels in the enlarged
scale. 
}
\label{lev1}
\end{figure}
  
We have shown, on the other hand,  \cite{damski01}
 that sweeping the laser beam
across the condensate along the spiral line may serve as an efficient
and stable way to create vortices in the system. 
This can be
easily understood by looking at the energy levels of the
Hamiltonian (\ref{ham}) for different 
(fixed) values of $x_0$, see Fig.~\ref{lev1}. 
Narrow avoided crossings
between neighboring energy levels indicate that starting with 
the system in the
ground state and changing $x_0$ from some negative
 value to zero, one may pass
 the avoided crossing diabatically and
end up (with a high efficiency) in the first excited state of the trap
that possesses the angular momentum $L_z=1$. 
Moreover, the process may be repeated,
Fig. ~\ref{lev1} suggests that having the system in the first 
excited state of the harmonic trap with $L_z=1$ (after a single sweep)
it is possible to employ the second similar process
and transfer the population to $L_z=2$ state with a high
efficiency.

\begin{figure}
\centering
{
\includegraphics[width=8.0cm, clip=true]{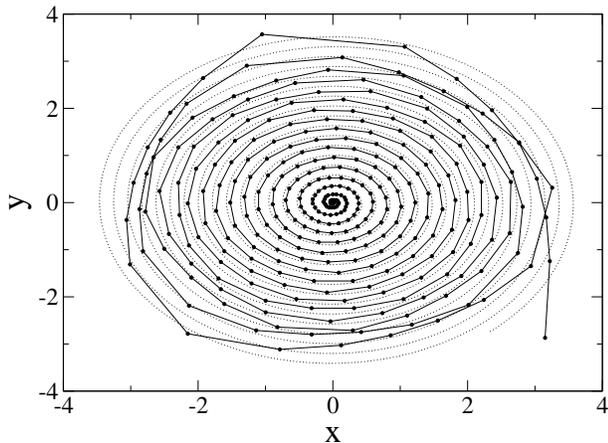}
}
\caption{ Trajectories (in the laboratory frame) of the vortex 
with $n=1$ (solid line)
and of the center of the laser beam (dotted line) from $t=145$ up to 
the end of the potential sweeping ($t=325$). The ground state of the harmonic
oscillator was chosen as an initial state and the parameters of the laser beam
were $u_0=16$, $\sigma=0.2$, $x_0(0)=-6.5$, $\Omega=0.6$  and $v=0.02$.
The positions of the vortex (circles on the solid line) 
were taken with time step that equals $0.54$.}
\label{vor1}
\end{figure}
To look more quantitatively at the stirring process,
we write the wave function of the system, in the hydrodynamical approach
\cite{ghosh}, as 
\protect{$\Psi(\vec{r}, t)=\sqrt{\rho(\vec{r}, t)} \exp(i \chi(\vec{r}, t))$}, 
where $\rho(\vec{r}, t)$ is the density of a probability fluid.
The velocity field is defined as  
\begin{equation}
\label{velocity}
\vec{v} =  \vec{\nabla}\chi(\vec{r}, t).
\end{equation}
The single valuedness of the wave function requires that 
the circulation 
of the velocity field $\Gamma_C$ around any closed contour $C$ is 
quantized (Feynman-Onsager quantization condition \cite{feyn})
\begin{equation}
\label{circ}
\Gamma_C = \oint_{C} \vec{v} \cdot d\vec{l} =2\pi  n,
\end{equation}
where  $n = 0, \pm 1, \pm 2, \dots$. 
The value of $n$ characterizes vortices in the wave function. We say that we
have vortex with unit charge at a given point, when calculation of $\Gamma_C$
gives $n=1$ as contour $C$, encircling that point, shrinks down to this point.
 
Creation of vortices by our method, which is nothing but a smooth
time-dependent modification of the potential,  requires a sudden
(due to the quantization) appearance 
of a non-zero circulation. This is necessarily accompanied by an
appearance of a singularity in the
velocity field. It is interesting to find out how this process occurs  since
we know that at the beginning of the laser sweep  there is
no circulation in the velocity field, at the end there is a vortex 
approximately at the center of the trap.  
Integrating  the time-dependent Schr\"odinger equation
(recall that we discuss noninteracting particles first) 
we have looked for  the wave function modulus minimum and calculated
the circulation around small contour encircling it. 
If $n$ is equal to $1 \pm 0.04$ ($-1 \pm 0.04$) we assume that vortex
(antivortex)
with unit charge is located at such a minimum.
As almost non-interacting 
condensates have been realized in laboratories already \cite{khyakovich,cornel3},
 it is perfectly legitimate to consider non-interacting particle case first.

\begin{figure}
\centering
{
\includegraphics[width=8.0cm, clip=true]{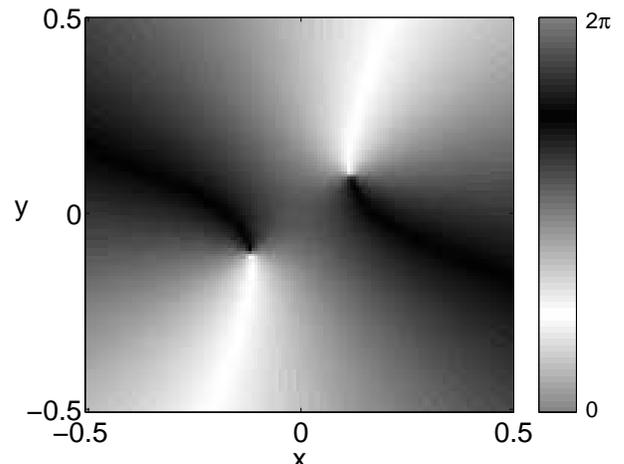}}
\caption{Plot of the phase of the final 
wave function after the potential sweeping with the laser beam 
parameters $u_0=16$,
$\sigma=0.2$, $x_0(0)=-6.5$, $\Omega=0.6$ and $v=0.02$. 
The eigenstate of the harmonic oscillator with $L_z=1$ was chosen as an initial
state. Despite the fact that the square overlap of the final wave 
function on the $L_z=2$ eigenstate is very high , 
there is not a single vortex
with the topological charge $n=2$ but two separate
vortices with $n=1$ -- see text.
}
\label{dwa}
\end{figure}
Let us inspect the first sweep of the laser beam through the system initially
in the ground state. 
We have found  (compare Fig.~\ref{vor1}) that the vortex
  moves in
from the {\it border} of the trap (i.e. the range of 
the configuration space where we are able to control 
the velocity field numerically). 
Please note that we are able to observe the vortex after 
some time since the beginning of the simulation.
 Indeed, it crosses
the trap {\it border} instead of being suddenly created at $t=145$---see Fig.~\ref{vor1}.
The position
of the vortex follows quite
closely the center of the focused laser beam. At the end of the excitation
process the vortex lands very close to the trap center.

Similarly for a second laser sweep aiming at increasing $L_z$ to 2
 an additional 
vortex with the topological charge $n=1$
comes from the  border of the trap 
along a spiral line (similar to the line depicted in
Fig.~\ref{vor1}) and collides with the first vortex which,
during the whole time evolution, is situated in the vicinity of 
the trap center.
In the numerical implementation the final wave function consists mainly of
the eigenstate with $L_z=2$ (the square overlap on this state is
$p_2\approx0.9997$). However, there is also a slight contribution from the 
$L_z=0$ eigenstate ($p_0\approx0.0003$). 
A simple calculation 
immediately shows 
that instead of a single vortex with $n=2$ we get two 
separate vortices with $n=1$ in this case.
This observation confirms that vortex with $n=2$ is unstable.
The two vortices are
situated symmetrically with respect to 
the trap center at a distance $2 (2 p_0/p_2)^{1/4}$. Plot of the phase of
the final wave function in the vicinity of the trap center 
confirms such  prediction, see Fig.~\ref{dwa}.

\begin{figure}
\centering
{
\includegraphics[width=8.0cm, clip=true]{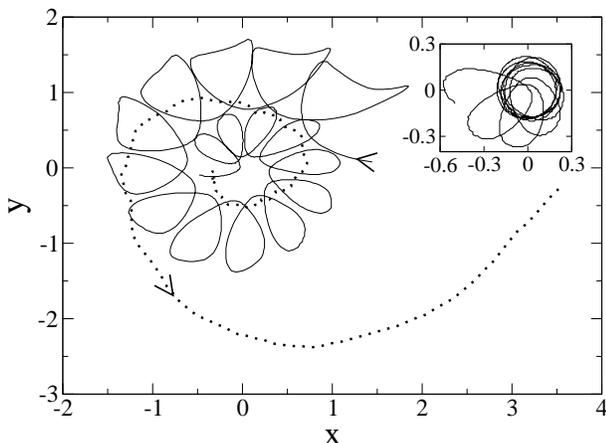}}

\caption{ Trajectories of vortices with the topological charge $n=1$ 
(dotted line) and $n=-1$ (solid line) during the potential sweeping. 
The $L_z=1$ eigenstate of the harmonic oscillator 
was chosen as an initial state and the parameters of the laser beam
were $u_0=16$, $\sigma=0.2$, $x_0(0)=-6.5$, $\Omega=0.25$  and $v=0.02$.
The main plot corresponds to $t\in [253,289]$.
After that time the vortex with
$n=1$ topological charge reaches the {\it border} of the trap and
further  evolution of the vortex 
with $n=-1$
up to the end of the potential sweeping ($t=325$) is shown in the inset.
The trajectory of this vortex 
ends a little off center at $(-0.035, -0.175)$.
}
\label{vor_1}
\end{figure}
Energy levels of the Hamiltonian (\ref{ham}) as a function of $x_0$ have been
calculated in Fig.~\ref{lev1} for $\Omega=0.6$. For $x_0=0$ 
the ground state corresponds to 
$L_z=0$, the first excited state corresponds to $L_z=1$ and the 
second one to 
$L_z=2$. However, the order can be
different if we decrease $\Omega$. Indeed for $\Omega<1/3$ the second
excited state (for $x_0=0$) corresponds to $L_z=-1$.
It offers an opportunity for the following counterintuitive situation 
which is of main interest for our study.
Suppose, we start with the $L_z=0$ state. After a potential sweeping
we end up with a very high probability in the state with $L_z=1$
where the rotation of the probability fluid coincides with the
rotation of the applied laser beam. Then another, identical stirring by our 
``laser spoon''  results in probability fluid rotating in the 
opposite direction (a state with $L_z=-1$)! Needless to say such a situation
is quite surprising and no analogy to some process  in a classical
fluid appears.

The prediction based on Hamiltonian  levels behavior can again be tested
by a direct integration of the time-dependent Schr\"odinger equation
and indeed the $L_z=-1$ state is excited with very high accuracy. 
Analyzing the process of such change of the angular momentum from 
$L_z=1$ to $L_z=-1$ by looking at the time dependent motion
of vortices we find that the vortex with $n=1$ initially situated at the
center moves out to the  {\it border} of the trap while the other vortex 
(born at  the  {\it border}) 
with an opposite $n=-1$ circulation 
arrives at the center along a complex trajectory shown in 
Fig.~\ref{vor_1}. The latter vortex, before reaching the center,	
experiences a sequence of collisions with 
another $n=1$ vortex that affects its trajectory. 
Therefore a transition from $n=1$ to $n=-1$ 
case is a result of (a bit complicated as seen in Fig.~\ref{vor_1}) dynamics
of vortices.

It remains to be seen whether the counterintuitive stirring scheme
is feasible  also in the presence of atom-atom interactions since 
so far we have presented a creation of vortices for a non-interacting BEC.
It is known, however, that the stability of vortices may be strongly affected
by  the atom-atom interactions \cite{ripoll}. 
To analyze the effect of interactions
we have performed numerical integration of the Gross-Pitaevskii 
equation \cite{pitaevskii}
\begin{equation}
\label{gpe}
i\frac{\partial \Psi}{\partial t}=(\hat{H} + g|\Psi|^2)\Psi,
\end{equation}
with $\hat{H}$ given by (\ref{ham}).
The interaction parameter $g$ is proportional to the number of atoms in the
system and to the $s$-wave scattering length. In an experiment, 
$g$ can be easily of order of thousands but it can be also reduced to a 
very small 
value exploring Feshbach resonances \cite{khyakovich,cornel3}.  
In the present work, we have chosen $g=100$ for the numerical calculations.
 
If the ground state of the system is chosen as an initial state,
applying the potential sweeping allows one to obtain the $L_z=1$
state with a high efficiency as described in Ref.~\cite{damski01}.
We performed such  numerical simulation taking $\Omega=0.1$.
Now, we apply the second similar laser sweep on the state obtained after the
first one.
It creates a vortex with the topological charge 
$n=-1$ similarly as it takes place for a noninteracting BEC  if 
$\Omega<1/3$. However, contrary to the linear case, the initial 
vortex with $n=1$ does not disappear 
--- the interaction between atoms makes the initial vortex more
robust to the perturbation. The vortex with $n=-1$ lands close
to the center of the trap while the original $n=1$ moves to the
edge of the trap.  In effect the total angular momentum per
particle is 
$\langle\hat{L}_z\rangle=-0.42$ with the dispersion
$\sigma_L=\sqrt{\langle\hat{L}^2_z\rangle-\langle\hat{L}_z\rangle^2}=1.13$.

\begin{figure}
\includegraphics[width=8.2cm, clip=true]{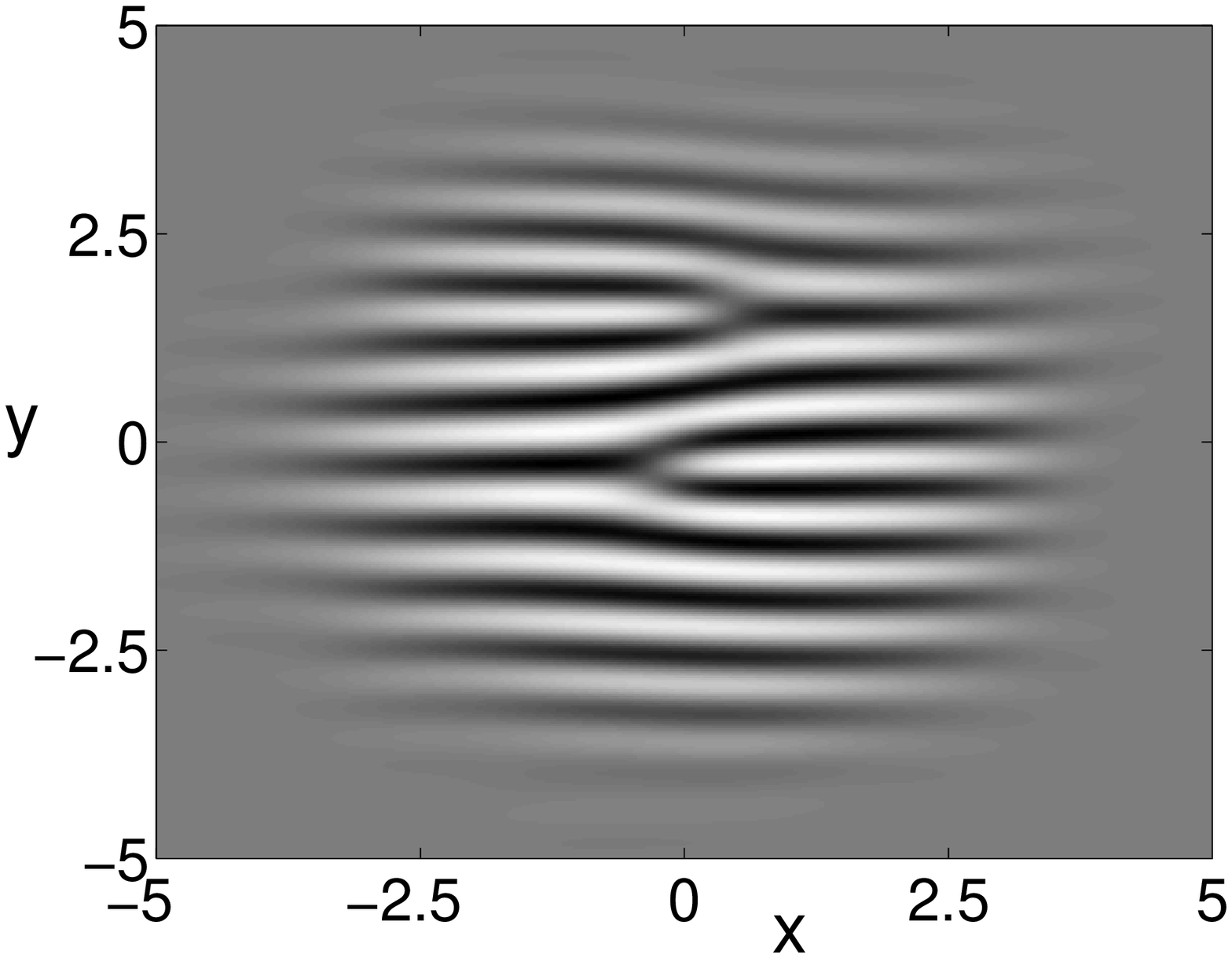}
\includegraphics[width=8.2cm, clip=true]{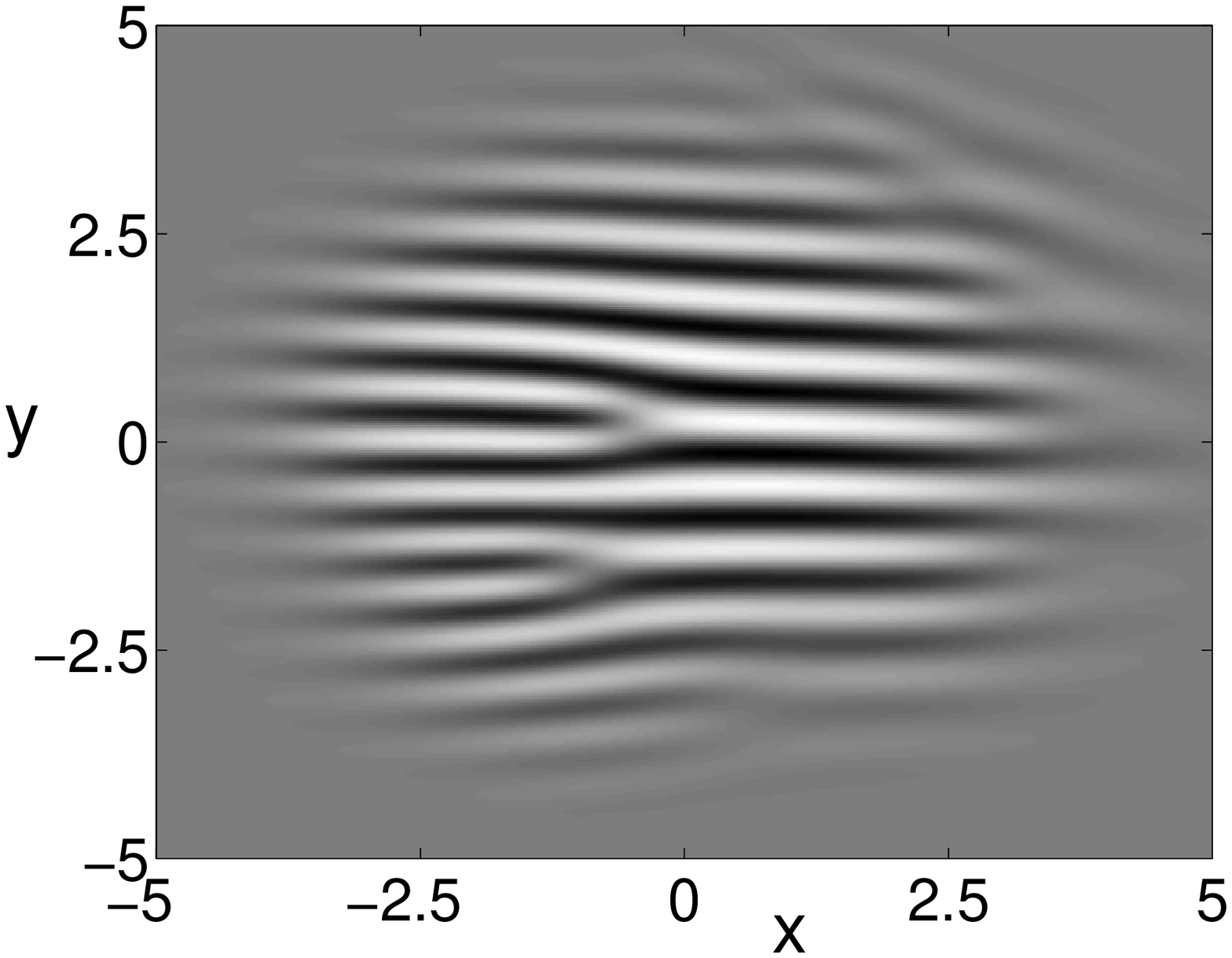} 
\caption{Interference pictures. Left: 
a well separated vortex - antivortex pair obtained after a
``second'' laser sweep through the harmonic potential
for $\Omega=0.1$ and the strength of the effective atom-atom
interaction $g=100$. For $\Omega=0.25$
one may observe two $n=1$ vortices (right). 
Atom-atom interaction ($g=100$)
leads to a big vortex separation - compare with Fig.~\protect{\ref{dwa}}. 
Parameters of the laser beam are $u_0=25$, $\sigma=0.2$, $x_0(0)=-6.5$, 
and $v=0.13$. Time of evolution was equal to 50.}
\label{widelce}
\end{figure}

The position of vortices may be observed using the interference approach
 \cite{bolda}.
In left frame of Fig.~\ref{widelce} the square modulus of the 
final wave function superimposed with a plane wave traveling 
vertically in the figure's plane is presented. 
A vortex--antivortex pair, clearly visible 
in the figure, can be observed experimentally as the interference technique
has been applied in a laboratory already \cite{kett}.
 The appearance of such a vortex-antivortex pair might be interesting from
an experimental point of view, since interactions between such pairs in BEC
   confined in a harmonic trap, are still an unexplored topic
 experimentally. 
 
\end{multicols}
\begin{multicols}{2}

Increasing the frequency $\Omega$ of the stirring to $\Omega=0.25$
we come
back to the case of two $n=1$ vortices discussed previously for noninteracting
case.  
Please note, that now the energy spacing between eigenstates with different 
value $L_z$ has decreased, so it is possible to have a $L_z=2$ state  as a
second excited  eigenstate for $\Omega<1/3$.
In the presence of atom-atom interaction ($g=100$) we again sweep
the laser across the condensate twice, first stirring creates a single
vortex, a second stirring process adds 
 an additional vortex with the topological 
charge $n=1$. This is again in a qualitative agreement with the 
noninteracting case considered previously. 
Quantitatively, the final state is
characterized by $\langle\hat{L}_z\rangle=1.69$ with $\sigma_L=1.92$.
The interaction between atoms leads now 
to a much larger separation between the two vortices,
 see the right frame in Fig.~\ref{widelce}. 
Indeed, the separation between them is now comparable with the size of 
the entire condensate (note different scales in Fig. ~\ref{dwa} and
Fig.~\ref{widelce}).

It is interesting to ask what is the 
critical stirring frequency for a transition from the regime of
`vortex-antivortex' to that of `vortex-vortex'  production
during  the second laser sweep.
 We estimate the critical frequency $\omega_c$ as satisfying
the following equation:
\[
\mu(L_z=-1)+\omega_c=\mu(L_z=2)-2\omega_c
\label{estima}
\] 
where $\mu(L_z=-1)$ and $\mu(L_z=2)$ are chemical potentials of two lowest 
eigenstates of the time-independent GP equation. 
The latter are found solving
the  2D 
equation $(\hat{H}+g|\Psi|^2)\Psi= \mu \Psi$, with Hamiltonian
$\hat{H}  =  -\frac{1}{2} \vec{\nabla}^2+ \frac{1}{2}(x^2 + y^2)$, i.e.,
the Hamiltonian ~(\ref{ham}) in the laboratory frame without laser beam.
For stirring frequencies  $\Omega$  lower than 
$\omega_c$ three lowest GP eigenstates,  in the frame rotating with stirrer,
possess angular momentum  
$L_z=0, 1, -1$ while  in the case of frequencies higher 
(but not too high) than $\omega_c$ the
order is  $L_z=0, 1, 2$.
The frequency $\omega_c$ is an {\it upper}
bound for the real critical frequency since its definition is based
solely on the ordering of eigenstates in the frame rotating with
stirrer. Indeed, an efficient transfer requires
 also that the distance in energy between
the level that we would like to populate and the next one should be sufficient
to assure adiabaticity, which is by definition not the case when we
stir the BEC with  $\Omega=\omega_c$. 
Therefore, one might expect that the optimal realization
 should require lower frequency, 
probably in the middle between $\omega_c$ and the lowest
estimate for a creation of vortex-antivortex pairs (equal to $0$).
That gives $\omega_c/2$ as a good guess.
We expect that the critical frequency should be somewhere between these two
estimates, namely between $\omega_c/2$ and $\omega_c$.
Calculation for $g=100$
gives $\omega_c=0.18$ which interestingly compares with  $0.125\pm 0.01$
 determined 
from a direct integration of Eq.~(\ref{gpe}) for  
$g=100$, $20\leq u_0 \leq25$ and a duration of a single laser sweep
between $40$ and $60$ (compare Fig.~\ref{widelce} and Eq.~(\ref{ham})).
Therefore, a numerical calculation
gives a value which is higher than
$\omega_c/2$ and lower than $\omega_c$, even though the lower bound is just a
rough estimate.
Similar  calculations of upper bounds for the critical frequency 
($\omega_c$) yield $0.237$ for $g=30$ and $0.195$ for $g=70$.

Finally, we would like to comment on an influence of the stirring scheme's
details on final results in the interacting case. 
First of all, we have observed 
that the width $\sigma$ [compare (\ref{ham})]
should be small, of the order of $0.2$; two times bigger
widths lead to a significant decrease in the stirring
process' efficiency. Secondly, the parameter $u_0$ 
 has to be high enough,
of the order of $20$, for an efficient  transfer of atoms from the 
ground state to the vortex
state(s). These two conditions provide non-trivial restrictions on laser beam
width and intensity, respectively.  Thirdly, changes of the switching time
 within about $20\%$ of a given time scale ($\approx 8$
periods of harmonic trap for $g=100$) do not affect the
dynamics qualitatively. Further details can be found in ~\cite{damski01}.

To summarize we have investigated the details of the creation  of 
vortices in BEC 
 when the laser sweep scheme \cite{damski01} is applied. 
Especially, we have shown that rotating the probability 
fluid by means of the ``laser spoon'' 
may introduce a circulation with the opposite direction with respect to 
the steering one. 

We are grateful to an anonymous referee for bringing references
 \cite{castin,caradoc99} to our attention and several useful hints.
Support of KBN under projects 5~P03B~088~21 (K.S. and J.Z.), 
2~P03B~124~22 (B.D.) is acknowledged. B. D. is grateful for hospitality 
extended to him in Hannover during preparation of the final form of
manuscript.

\end{multicols}

\end{document}